%% file: main.tex
\documentclass[%
 reprint,
 superscriptaddress,
 amsmath,amssymb,
 aps,
 prl,
 twocolumn
]{revtex4-2}

\usepackage{graphicx}
\usepackage{dcolumn}
\usepackage{amsmath}
\usepackage{bm}
\usepackage{colortbl}
\usepackage{float}
\usepackage{hyperref}
\usepackage[table,dvipsnames]{xcolor}
\usepackage{makecell}
\usepackage[mathscr]{euscript}
\usepackage{multirow}
\usepackage{braket}
\usepackage[pagewise]{lineno}
\usepackage{enumitem}
\usepackage{supertabular}
\usepackage{amssymb}
\usepackage{makecell}
\usepackage{tikz}
\usepackage{bbding}
\begin{document}

\title{Electrically switchable non-relativistic Zeeman spin splittings \\
in collinear antiferromagnets}

\author{Longju Yu}
\affiliation{Key Laboratory of Material Simulation Methods and Software of Ministry of Education, College of Physics, Jilin University, Changchun 130012, China}

\author{Hong Jian Zhao}
\affiliation{Key Laboratory of Material Simulation Methods and Software of Ministry of Education, College of Physics, Jilin University, Changchun 130012, China}
\affiliation{Key Laboratory of Physics and Technology for Advanced Batteries (Ministry of Education), College of Physics, Jilin University, Changchun 130012, China}
\affiliation{State Key Laboratory of High Pressure and Superhard Materials, College of Physics, Jilin University, Changchun 130012, China}
\affiliation{International Center of Future Science, Jilin University, Changchun 130012, China}

\author{Laurent Bellaiche}
 \affiliation{Smart Ferroic Materials Center, Physics Department and Institute for Nanoscience and Engineering, University of Arkansas, Fayetteville, Arkansas 72701, USA}
\affiliation{Department of Materials Science and Engineering, Tel Aviv University, Ramat Aviv, Tel Aviv 6997801, Israel}

\author{Yanming Ma}
\affiliation{School of Physics, Zhejiang University, Hangzhou 310058, China}
\affiliation{Key Laboratory of Material Simulation Methods and Software of Ministry of Education, College of Physics, Jilin University, Changchun 130012, China}
\affiliation{State Key Laboratory of High Pressure and Superhard Materials, College of Physics, Jilin University, Changchun 130012, China}
\affiliation{International Center of Future Science, Jilin University, Changchun 130012, China}

\begin{abstract}
Magnetic or electrical manipulation of electronic spin is elementary for spin-based logic, computing, and memory, where the latter is a low-power manipulation scheme. Rashba-like spin splittings stemming from spin-orbit interaction (SOI) enable electric-field manipulation of spin, but the relativistic SOI causes spin relaxations and yields dissipative transport of spin-encoded information. Recent works suggest the occurrence of electric-field switchable non-relativistic Zeeman spin splittings (NRZSSs) in collinear antiferromagnets --- allowing for electrical manipulation of spin in the non-relativistic regime; yet, a theory elucidating the mechanisms for these NRZSSs and guiding the materials discovery remains missing. Here, we develop such a theory by analyzing the symmetries of spin point groups. We highlight the linear magnetoelectric and bilinear piezomagnetoelectric mechanisms for NRZSSs that depend linearly on electric field and are electrically switchable. First-principles calculations further confirm that FeOOH and NaMnP showcase such NRZSSs. Our theory provides guidelines for discovering light-element collinear antiferromagnets with electrically switchable NRZSSs, which are promising for the design of high-performance spin-based devices.
\end{abstract}

\maketitle

\noindent
\textit{Introduction ---} 
Spintronic devices (e.g., logic, computing, and memory) are built upon the manipulation of electronic spin in solids~\cite{baltz2018antiferromagnetic,rimmler2024non,han2023coherent,he2022topological,lin2019two}. Such a manipulation can be achieved by magnetic field or electric field, where the latter offers fascinating low-power devices by reducing the ``Joule heating''~\cite{yan2020electric,fert2024electrical,manchon2015new}. In principle, electric-field manipulation of spin is enabled by electrically switchable spin splittings~\cite{tao2021perspectives,picozzi2014ferroelectric}. Because of the relativistic spin-orbit interaction (SOI), dielectrics and ferroelectrics naturally host electrically switchable Rashba-like spin splittings~\cite{tao2025ferroelectric,varignon2019electrically,noel2020non,sheng2021rashba,bihlmayer2022rashba}. As a by-product, SOI causes the so-called D'yakonov-Perel' spin relaxation as well~\cite{dyakonov1972spin,manchon2019current,wu2010spin}, which yields the spin decoherence of electrons and prevents the long-range transport of the spin-encoded information~\cite{maekawa2017spin,wu2010spin,schapers2021semiconductor,gonzalez2021efficient}. To suppress the D'yakonov-Perel' spin relaxation, the proposal of electrically switchable spin splittings in the non-relativistic regime is of high necessity.  

Recent works highlight two types of notable spin splittings: alternating spin splittings~\cite{hayami2019momentum,vsmejkal2022beyond,vsmejkal2022emerging,song2025altermagnets,zhang2025electrical,PhysRevLett.134.106802,vsmejkal2024altermagnetic,zhou2024transition} and non-relativistic Zeeman spin splittings (NRZSSs)~\cite{mazin2024altermagnetism,finley2020spintronics,zhou2021efficient,PhysRevB.108.L180403,PhysRevLett.121.077701,yuan2024nonrelativistic,gong2018electrically,zhao2022zeeman,zhang2024predictable,sheoran2024nonrelativistic,tao2024layer,sheoran2024multiple,yuan2023uncovering,liu2025two,chen2025gate}, both of which provide platforms for achieving electrical manipulation of spin within the non-relativistic regime. Alternating spin splittings occur in altermagnets, which can exhibit $d$-, $g$-, or $i$-wave anisotropy in the Brillouin zone~\cite{hayami2019momentum,vsmejkal2022beyond,vsmejkal2022emerging,song2025altermagnets}, and in certain cases, such splittings can be controlled by electric fields~\cite{PhysRevLett.134.106802,vsmejkal2024altermagnetic,zhou2024transition}. Unlike the alternating spin splittings, NRZSSs lift spin degeneracy uniformly across the entire Brillouin zone~\cite{yuan2024nonrelativistic,liu2025two}. This feature enables the generation of fully spin-polarized currents~\cite{PhysRevLett.74.1171,yuan2024nonrelativistic,liu2025two}, which is ideal for highly efficient spin injection in spintronic devices~\cite{wolf2001spintronics,xie2025emerging}. Notably, theoretical works suggest the possibility of achieving NRZSSs in collinear antiferromagnets that are switchable by an electric field~\cite{gong2018electrically,zhao2022zeeman,zhang2024predictable,liu2025two,sheoran2024nonrelativistic,tao2024layer,sheoran2024multiple,yuan2023uncovering,chen2025gate}. For example, electrically switchable NRZSSs were predicted in two-dimensional layer materials (e.g., bilayer FeBr$_2$~\cite{yuan2023uncovering} and bilayer VSe$_2$~\cite{gong2018electrically}) and bulk materials (e.g., Fe$_2$TeO$_6$~\cite{zhao2022zeeman}). Regarding electrically switchable Zeeman spin splittings, a theory was developed by analyzing the magnetic point groups (involving the SOI)~\cite{zhao2022zeeman}. Such a theory, rooted in the relativistic regime, can neither interpret the aforementioned NRZSSs nor predict materials with electrically switchable NRZSSs.

So far, mechanisms for electric-field switchable NRZSSs are elusive and selection rules for the corresponding materials discovery are missing. To address this issue, we analyze the symmetries of spin point groups (SPGs) that are associated with collinear magnets in the non-relativistic limit. This allows us to develop a theory on electrically switchable NRZSSs in collinear antiferromagnets, and reveal the magnetoelectric and piezomagnetoelectric mechanisms for this phenomenon. Our theory provides theoretical guidelines for screening collinear antiferromagnets with electrically switchable NRZSSs. This yields the first-principles predictions that FeOOH and NaMnP are two candidates with linear magnetoelectric NRZSSs and bilinear piezomagnetoelectric NRZSSs, respectively.\\

\noindent
\textit{NRZSSs in collinear magnets ---} Without the inclusion of SOI, the electronic spin angular momentum is conserved in collinear magnets~\footnote{To design spin-based devices, spin conservation is a desired feature~\cite{maekawa2017spin}. While spin conservation is a merit of non-relativistic collinear magnets, the occurrence of spin conservation is not guaranteed in non-relativistic noncollinear magnets (see e.g., Refs.~\cite{ vsmejkal2022NRM,yuan2021prediction, vsmejkal2022beyond}). The non-relativistic noncollinear magnets are thus not of interest to us.} and serves as a good quantum number to label the electronic eigen states therein~\cite{vsmejkal2022NRM,yuan2021prediction, vsmejkal2022beyond, vsmejkal2022emerging}. In this case, each spin- and momentum-resolved eigen state is associated with an eigen energy of $\varepsilon(\mathbf{k},s_\chi)$, where $\mathbf{k}$ is the electronic momentum, $s_\chi$ the spin angular momentum, and $\chi$ the characteristic direction for collinear magnets (i.e., magnetic moments being along $\boldsymbol{\chi}$ or $-\boldsymbol{\chi}$). 
According to Refs.~\cite{liu2022spin, chen2024enumeration, litvin1974spin, litvin1977spin}, 
the symmetries of non-relativistic collinear magnets can be well described by 32 ferromagnetic or ferrimagnetic SPGs and 58 antiferromagnetic SPGs. The symmetry operations in SPGs are represented by $[R_s \left| \right| R_l]$, with the spin operation $R_s$ and spatial operation $R_l$ being separated by the ``$\left| \right|$'' symbol~\cite{ liu2022spin, chen2024enumeration}. The $[R_s \left| \right| R_l]$ symmetry operation links the $(\mathbf{k},s_\chi)$ and $(R_lR_s\mathbf{k},R_s s_\chi)$ eigen states~\footnote{The spin operation can be written as $R_s=U_s$ or $R_s=U_sT$, with $U_s$ and $T$ being spin proper rotation and time reversal operation, respectively. According to Refs.~\cite{liu2022spin, chen2024enumeration}, $U_s\mathbf{k}=\mathbf{k}$ and $U_sT\mathbf{k}=-\mathbf{k}$.}, implying the $\varepsilon(\mathbf{k},s_\chi)=\varepsilon(R_lR_s\mathbf{k},R_s s_\chi)$ relation. At the Brillouin zone center, both $\mathbf{k}=0$ and $R_lR_s\mathbf{k}=0$ are valid so that the aforementioned relation can be simplified as $\varepsilon(s_\chi)=\varepsilon(R_s s_\chi)$ by omitting $\mathbf{k}$.

For all the $[R_s \left| \right| R_l]$ symmetry operations in ferromagnetic or ferrimagnetic SPGs, $R_s s_\chi=s_\chi$ is always true and $\varepsilon(s_\chi)=\varepsilon(-s_\chi)$ is not necessarily enforced~\cite{liu2022spin, chen2024enumeration}. This means that non-relativistic collinear ferromagnets or ferrimagnets enable the spontaneous NRZSSs (see e.g., Refs.~\cite{yuan2021prediction, vsmejkal2022beyond, vsmejkal2022emerging}). Unlike those in ferromagnetic or ferrimagnetic SPGs, the $[R_s \left| \right| R_l]$ symmetry operations in an antiferromagnetic SPG can be partitioned into two subsets~\cite{liu2022spin, chen2024enumeration, vsmejkal2022beyond, vsmejkal2022emerging}, namely,
\begin{equation}\label{eq:L+}
    L_{+}=\{ [R_s^+ \left| \right| R_l^+]: R_s^+ s_\chi=s_\chi \},
\end{equation}
and
\begin{equation}\label{eq:L-}
    L_{-}=\{ [R_s^- \left| \right| R_l^-]: R_s^- s_\chi=-s_\chi \}.
\end{equation}
The $L_{+}$ subset is a subgroup of the antiferromagnetic SPG, and the symmetry operations therein have no restriction on $\varepsilon(s_\chi)$. In contrast, the symmetry operations in $L_{-}$ suggest the $\varepsilon(s_\chi)=\varepsilon(-s_\chi)$ degeneracy, and prevent the spontaneous NRZSSs~\cite{yuan2021prediction, vsmejkal2022beyond, vsmejkal2022emerging}. On the other hand, the application of external stimuli may induce NRZSSs in non-relativistic collinear antiferromagnets. This is achieved by that such stimuli break \textit{all} the symmetry operations in the $L_{-}$ set. \\

\noindent
\textit{Electric-field induced NRZSSs ---} We concentrate on the NRZSSs in nonpolar collinear antiferromagnets that are driven by external electric field. To illustrate our basic logic, we work with an antiferromagnet whose SPG is the union of $L_{+}$ and $L_{-}$ sets [see Eqs.~(\ref{eq:L+}) and~(\ref{eq:L-})]. The application of electric field $E_\alpha$ along $\alpha$ direction breaks the symmetries of the antiferromagnet ($\alpha$ being $x$, $y$, or $z$). This reduces the SPG of the antiferromagnet to the union of $L_{+}^\alpha$ and $L_{-}^\alpha$ sets, where 
\begin{equation}\label{eq:Lreduc}
\begin{split}
    L_{+}^\alpha =\{ [R_s^+ \left| \right| R_{l}^+]\in L_+:  R_{l}^+ E_\alpha = E_\alpha \}, \\
    L_{-}^\alpha =\{ [R_s^- \left| \right| R_{l}^-]\in L_-:  R_{l}^- E_\alpha = E_\alpha \},        
\end{split}
\end{equation}
and the ``$\alpha$'' superscript labels the symmetry reduction caused by the $E_\alpha$ electric field~\footnote{When $E_\alpha$ is not compatible with the spatial $R_l$ operation (i.e., $R_l E_\alpha \neq E_\alpha$), the application of $E_\alpha$ break the $[R_s \left| \right| R_l]$ symmetry operation.}. When $L_{-}^\alpha$ becomes empty, there will be no symmetry operations that enforce the $\varepsilon(s_\chi)=\varepsilon(-s_\chi)$ degeneracy --- enabling the electric-field induced NRZSSs. The $E_\alpha$-field-induced NRZSSs are described by the effective Hamiltonian
\begin{equation}\label{eq:HME}
\begin{split}
\mathcal{H}= \lambda_1 E_{\alpha}\sigma_{\chi} + \lambda_2 E_{\alpha}^2\sigma_{\chi} +  \cdots, 
\end{split}
\end{equation}
with $\lambda_n$ ($n=1,2,\cdots$) characterizing the 
$n_\mathrm{th}$-order response of the Zeeman splitting to the $E_\alpha$ field, and $\sigma_{\chi}$ representing the Pauli matrix along the $\chi$ direction. Eq.~(\ref{eq:HME}) is reminiscent of the magnetoelectric coupling, and this motivates us to name the electric-field induced NRZSSs as the magnetoelectric NRZSSs.

The $L_{-}^\alpha$ set associated with an SPG may be not empty, and a material with this SPG does not host the $E_\alpha$-induced NRZSSs. In such a case, we can grow this material on an appropriate substrate so that an epitaxial $\eta_{\beta\gamma}$ strain is exerted to this material. The $\eta_{\beta\gamma}$ strain might modify the symmetry of the material, and reduce the $L_{-}$ set to $L_{-}^{\beta\gamma} =\{ [R_s^- \left| \right| R_{l}^-] \in L_{-}: R_{l}^- \eta_{\beta\gamma} = \eta_{\beta\gamma} \}$. Applying the $E_\alpha$ electric field to the strained material further reduces $L_{-}^{\beta\gamma}$ to
\begin{equation}\label{eq:LLreduc}
    L_{-}^{\alpha\beta\gamma} =\{ [R_s^- \left| \right| R_{l}^-] \in L_{-}^{\beta\gamma}:R_{l}^- E_\alpha = E_\alpha \}.        
\end{equation}
Provided that $L_{-}^{\alpha\beta\gamma}$ is empty but $L_{-}^{\beta\gamma}$ is not, the NRZSSs are created via the cooperation of $E_\alpha$ and $\eta_{\beta\gamma}$~\footnote{When $L_{-}^{\beta\gamma}$ is empty, the NRZSSs can be driven by the $\eta_{\beta\gamma}$ strain solely, without involving the $E_\alpha$ electric field.}. Phenomenologically, such NRZSSs are described by the effective Hamiltonian
\begin{equation}\label{eq:HMEeta}
\mathcal{H}^\prime= \lambda^\prime_1 E_{\alpha}\eta_{\beta\gamma}\sigma_{\chi} + \lambda^\prime_2 E_{\alpha}^2\eta_{\beta\gamma}\sigma_{\chi} + \lambda^\prime_3 E_{\alpha}\eta_{\beta\gamma}^2\sigma_{\chi}+ \cdots, 
\end{equation}
with $\lambda^\prime_n$ ($n=1,2,\cdots$) being the coupling coefficients. Eq.~(\ref{eq:HMEeta}) motivates us to name this type of NRZSSs as piezomagnetoelectric NRZSSs.\\

\begin{table}[!ht]
\centering
\renewcommand{\arraystretch}{1.45}
\setlength{\tabcolsep}{0.5mm}{\caption{\label{tab:ME} {The nonpolar collinear SPGs that enable linear magnetoelectric NRZSSs. 
For an SPG associated with the $E_\alpha\sigma_\chi$ coupling, a ``$\circ$'' symbol is filled in the corresponding entry; Such an entry is endowed with a ``$-$'' symbol if the $E_\alpha\sigma_\chi$ coupling is not symmetrically allowed. For short, we relabel each collinear $G^{\infty m}1$ SPG as $G$ by omitting its $^{\infty m}1$ part. As an example, the $^{1}m^{1}m^{\bar{1}}m^{\infty m}1$ SPG (abbreviated as $^{1}m^{1}m^{\bar{1}}m$) hosts the $E_z$-driven linear magnetoelectric NRZSSs via the $E_z\sigma_\chi$ coupling, but does not accommodate the $E_x$- or $E_y$-driven linear magnetoelectric NRZSSs. 
}}
\begin{tabular}{b{1.76cm}<{\raggedright}b{0.7cm}<{\centering}b{0.7cm}<{\centering}b{0.7cm}<{\centering}|b{1.76cm}<{\raggedright}b{0.7cm}<{\centering}b{0.7cm}<{\centering}b{0.7cm}<{\centering}}
\toprule
SPGs & $E_x$ & $E_y$ & $E_z$ &
SPGs & $E_x$ & $E_y$ & $E_z$ \\
\hline
$^{\bar{1}}\bar{1}$  & $\circ$ &$\circ$&$\circ$&
$^{\bar{1}}2/^{1}m$  & $\circ$ & $-$ & $\circ$ \\
\hline
$^{1}2/^{\bar{1}}m$  & $-$&$\circ$&$-$ &
$^{\bar{1}}2^{\bar{1}}2^{1}2$  & $-$& $-$ &$\circ$ \\
\hline
$^{\bar{1}}2^{1}2^{\bar{1}}2$ & $-$ &$\circ$& $-$ & 
$^{1}2^{\bar{1}}2^{\bar{1}}2$ & $\circ$ & $-$ & $-$ \\
\hline
$^{\bar{1}}m^{1}m^{1}m$ & $\circ$ & $-$ & $-$ &
$^{1}m^{\bar{1}}m^{1}m$ & $-$ & $\circ$  & $-$ \\
\hline
$^{1}m^{1}m^{\bar{1}}m$ & $-$ & $-$ & $\circ$ & 
$^{\bar{1}}\bar{4}$ & $-$ & $-$ & $\circ$ \\
\hline
$^{1}4/^{\bar{1}}m$ & $-$ & $-$ & $\circ$ &
$^{1}4^{\bar{1}}2^{\bar{1}}2$ & $-$ & $-$ & $\circ$  \\
\hline
$^{\bar{1}}\bar{4}^{\bar{1}}2^{1}m$ & $-$ & $-$ & $\circ$ & 
$^{1}4/^{\bar{1}}m^{1}m^{1}m$ & $-$ & $-$ & $\circ$ \\
\hline
$^{\bar{1}}\bar{3}$ & $-$ & $-$ & $\circ$ & 
$^{1}3^{\bar{1}}2$ & $-$ & $-$ & $\circ$ \\
\hline
$^{\bar{1}}\bar{3}^{1}m$ & $-$ & $-$ & $\circ$ & 
$^{\bar{1}}\bar{6}$ & $-$ & $-$ & $\circ$ \\
\hline
$^{1}6/^{\bar{1}}m$ & $-$ & $-$ & $\circ$ &
$^{1}6^{\bar{1}}2^{\bar{1}}2$ & $-$ & $-$ & $\circ$ \\
\hline
$^{\bar{1}}\bar{6}^{1}m^{\bar{1}}2$ & $-$ & $-$ & $\circ$ & 
$^{1}6/^{\bar{1}}m^{1}m^{1}m$ & $-$ & $-$ & $\circ$ \\
 \hline
  \hline
\end{tabular}}
\end{table}

\noindent
\textit{Linear magnetoelectric and bilinear piezomagnetoelectric NRZSSs ---} We continue to carry out symmetry analysis and extract the SPGs that host the electric-field switchable NRZSSs, following the logics mentioned in the previous section. Our detailed analysis can be found in Sections I and II of the Supplementary Material (SM)~\footnote{See Supplemental Material at \url{[URL]} for the analysis of spin point groups, the analysis of electric-field induced non-relativistic Zeeman spin splittings, methods, some numerical results, and the analysis of FeOOH.}, which includes Refs.~\cite{ gtpack2,dresselhaus2007group, kresse1996efficient, kresse1999ultrasoft, blochl1994projector, ceperley1980ground, PhysRevB.57.1505, fu2003first,chen2019electric,chen2016giant, wang2021vaspkit, momma2011vesta, stokes2005findsym, herath2020pyprocar,LANG2024109063, aroyo2011crystallography, aroyo2006bilbao1, aroyo2006bilbao2, gallego2016magndata-I, gallego2016magndata-II,Hunter2007}. The SPGs hosting magnetoelectric and piezomagnetoelectric NRZSSs are summarized in Tables S6 and S9 of the SM.
As shown in Eqs.~(\ref{eq:HME}) and~(\ref{eq:HMEeta}), the $E_\alpha$-driven magnetoelectric or piezomagnetoelectric NRZSSs may be rooted in $E_\alpha \sigma_\chi$, $E_{\alpha}\eta_{\beta\gamma}\sigma_{\chi}$, $E_\alpha^2 \sigma_\chi$, 
$E_{\alpha}\eta_{\beta\gamma}^2\sigma_{\chi}$, or other higher-order couplings (see Tables S11 and S12 of the SM). Here, we focus on the predominant electric-field induced NRZSSs which include linear magnetoelectric NRZSSs and bilinear piezomagnetoelectric NRZSSs. In Tables~\ref{tab:ME} and~\ref{tab:ETAME}, we collect the nonpolar collinear SPGs that enable these two types of NRZSSs.

The linear magnetoelectric NRZSSs and bilinear piezomagnetoelectric NRZSSs are characterized by $\lambda_1 E_\alpha \sigma_\chi$ and $\lambda^\prime_1 E_{\alpha}\eta_{\beta\gamma}\sigma_{\chi}$, respectively. At the Brillouin zone center, the electronic eigen energies are $\varepsilon(\pm s_\chi) = \varepsilon_0 \pm \lambda_1 E_\alpha$ or $\varepsilon (\pm s_\chi) = \varepsilon_0 \pm \lambda^\prime_1 E_{\alpha}\eta_{\beta\gamma}$, with $\varepsilon_0$ being an energy independent of spin. When changing the electric field from $E_\alpha$ to $-E_\alpha$, the eigen energies become $\varepsilon(\pm s_\chi) = \varepsilon_0 \mp \lambda_1 E_\alpha$ or $\varepsilon (\pm s_\chi) = \varepsilon_0 \mp \lambda^\prime_1 E_{\alpha}\eta_{\beta\gamma}$. This means that both linear magnetoelectric NRZSSs and bilinear piezomagnetoelectric NRZSSs are electric-field switchable.\\

\begin{table}[!ht]
\centering
\renewcommand{\arraystretch}{1.4}
\setlength{\tabcolsep}{0.5mm}{\caption{\label{tab:ETAME} {
The nonpolar collinear SPGs that enable the bilinear piezomagnetoelectric NRZSSs. The $\eta_{xx}$, $\eta_{yy}$, $\eta_{zz}$, $\eta_{xy}$, $\eta_{yz}$, and $\eta_{zx}$ strains are represented by ``1'', ``2'', ``3'', ``4'', ``5'', and ``6'' numbers, respectively. For an SPG associated with the $E_\alpha\eta_{\beta\gamma}\sigma_{\chi}$ coupling, the number representing the $\eta_{\beta\gamma}$ strain is filled in the corresponding entry. If an SPG enables $E_\alpha\eta_{\beta\gamma}\sigma_{\chi}$ and $E_\alpha\eta_{\beta^\prime\gamma^\prime}\sigma_{\chi}$ couplings, two numbers representing the $\eta_{\beta\gamma}$ and $\eta_{\beta^\prime\gamma^\prime}$ strains are filled in the corresponding entry, and so forth. The entry with ``$-$'' indicates that the corresponding bilinear piezomagnetoelectric NRZSSs are symmetrically forbidden. In this table, we relabel each collinear $G^{\infty m}1$ SPG as $G$ by omitting its $^{\infty m}1$ part. As an example, the $^{\bar{1}}\bar{3}^{1}m^{\infty m}1$ SPG (abbreviated as $^{\bar{1}}\bar{3}^{1}m$ in this table) hosts the $E_x$-driven bilinear piezomagnetoelectric NRZSSs via $E_x\eta_{xy}\sigma_{\chi}$ or $E_x\eta_{zx}\sigma_{\chi}$ couplings, but does not accommodate the $E_y$- and $E_z$-driven bilinear piezomagnetoelectric NRZSSs. }}
\begin{tabular}{b{1.76cm}<{\raggedright}b{0.72cm}<{\centering}b{0.72cm}<{\centering}b{0.72cm}<{\centering}|b{1.76cm}<{\raggedright}b{0.72cm}<{\centering}b{0.72cm}<{\centering}b{0.72cm}<{\centering}}
\toprule
SPGs & $E_x$ & $E_y$ & $E_z$ & SPGs & $E_x$ & $E_y$ & $E_z$  \\
\hline
$^{\bar{1}}2/^{1}m$ & $-$ & $45$& $-$ & $^{1}2/^{\bar{1}}m$ & $45$ & $-$& $45$\\
\hline
$^{\bar{1}}2^{\bar{1}}2^{1}2$&$6$&$5$&$-$&$^{\bar{1}}2^{1}2^{\bar{1}}2$&$4$&$-$&$5$\\
\hline
$^{1}2^{\bar{1}}2^{\bar{1}}2$&$-$&$4$&$6$&$^{\bar{1}}m^{1}m^{1}m$&$-$&$4$&$6$\\
\hline
$^{1}m^{\bar{1}}m^{1}m$&$4$&$-$&$5$&$^{1}m^{1}m^{\bar{1}}m$&$6$&$5$&$-$\\
\hline
$^{\bar{1}}m^{\bar{1}}m^{\bar{1}}m$&$5$&$6$&$4$&$^{1}4/^{\bar{1}}m$&$56$&$56$&$-$\\
\hline
$^{\bar{1}}4/^{\bar{1}}m$&$56$&$56$&$124$&$^{\bar{1}}4^{1}2^{\bar{1}}2$&$-$&$-$&$4$\\
\hline
$^{\bar{1}}4^{\bar{1}}2^{1}2$&$6$&$5$&$12$&$^{1}4^{\bar{1}}2^{\bar{1}}2$&$6$&$5$&$-$\\
\hline
$^{\bar{1}}\bar{4}^{\bar{1}}2^{1}m$&$6$&$5$&$-$&$^{1}\bar{4}^{\bar{1}}2^{\bar{1}}m$&$6$&$5$&$12$\\
\hline
$^{1}4/^{\bar{1}}m^{1}m^{1}m$&$6$&$5$&$-$&$^{\bar{1}}4/^{\bar{1}}m^{\bar{1}}m^{1}m$&$5$&$6$&$4$\\
\hline
$^{\bar{1}}4/^{\bar{1}}m^{1}m^{\bar{1}}m$&$6$&$5$&$12$&$^{1}4/^{\bar{1}}m^{\bar{1}}m^{\bar{1}}m$&$5$&$6$&$-$\\
\hline
$^{\bar{1}}\bar{3}^{1}m$&$46$&$-$&$-$&
$^{\bar{1}}\bar{3}^{\bar{1}}m$&$-$&$46$&$-$\\
\hline
$^{1}6/^{\bar{1}}m$&$56$&$56$&$-$&
$^{\bar{1}}6^{\bar{1}}2^{1}2$&$4$&$-$&$-$\\
\hline
$^{\bar{1}}6^{1}2^{\bar{1}}2$&$-$&$4$&$-$&
$^{1}6^{\bar{1}}2^{\bar{1}}2$&$6$&$5$&$-$\\
\hline
$^{\bar{1}}\bar{6}^{\bar{1}}m^{1}2$&$5$&$-$&$-$&
$^{\bar{1}}\bar{6}^{1}m^{\bar{1}}2$&$6$&$-$&$-$\\
\hline
$^{1}6/^{\bar{1}}m^{1}m^{1}m$&$6$&$5$&$-$&
$^{\bar{1}}6/^{1}m^{\bar{1}}m^{1}m$&$-$&$4$&$-$\\
\hline
$^{\bar{1}}6/^{1}m^{1}m^{\bar{1}}m$&$4$&$-$&$-$&
$^{1}6/^{\bar{1}}m^{\bar{1}}m^{\bar{1}}m$&$5$&$6$&$-$\\
\hline
$^{\bar{1}}m^{\bar{1}}\bar{3}$&$5$&$6$&$4$&
$^{\bar{1}}4^{1}3^{\bar{1}}2$&$5$&$6$&$4$\\
\hline
$^{\bar{1}}m^{\bar{1}}\bar{3}^{1}m$ &$5$&$6$&$4$\\
 \hline
 \hline
\end{tabular}}
\end{table}

\noindent
\textit{Materials with electric-field switchable NRZSSs ---} 
We recall that the electrically switchable NRZSSs in bilayer FeBr$_2$, bilayer VSe$_2$, bulk Fe$_2$TeO$_6$ and other materials, previously predicted in Refs.~\cite{gong2018electrically,zhao2022zeeman,zhang2024predictable,liu2025two,sheoran2024nonrelativistic,tao2024layer,sheoran2024multiple,yuan2023uncovering,chen2025gate}, can be ascribed to the linear magnetoelectric mechanisms in our theory (see Section II.D of the SM). Next, we explore more materials that might showcase electric-field switchable NRZSSs (i.e., linear magnetoelectric or bilinear piezomagnetoelectric NRZSSs). This is assisted by the MAGNDATA database~\cite{gallego2016magndata-I,gallego2016magndata-II}, and guided by Tables~\ref{tab:ME} and~\ref{tab:ETAME}. We search for nonpolar collinear antiferromagnets composed of light elements, identify the SPGs for such antiferromagnets, and discover the FeOOH and NaMnP as two representatives with sizeable electric-field switchable NRZSSs. [The strength of SOI basically depends on the atomic mass. Materials composed of heavy elements (e.g., Bi, Sb, and Pt) likely showcase sizable SOI~\cite{zhang2009topological,li2022designing,liu2011spin}.] Especially, both FeOOH and NaMnP are room-temperature antiferromagnets, with N\'{e}el temperature being $>$343~K~\cite{zepeda2014magnetic,ozdemir1996thermoremanence} for FeOOH and $>$293~K~\cite{bronger1985ternary,bronger1986charakterisierung} for NaMnP.

\begin{figure}[!ht]
\includegraphics[width=1.\linewidth]{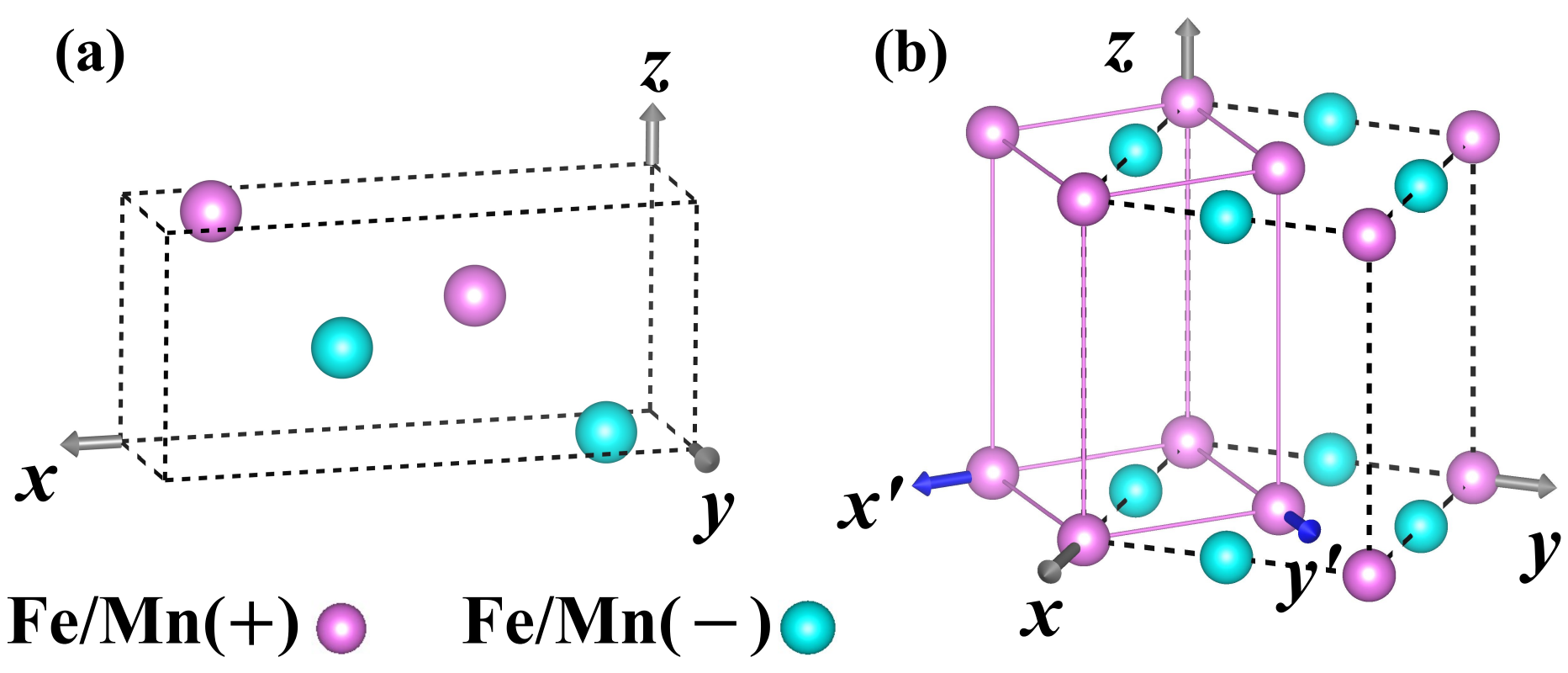}
\caption{\label{fig:stru} Collinear magnetic structures for (a) FeOOH and (b) NaMnP antiferromagnets, where only the Fe and Mn sublattices are displayed. The cyan and pink spheres denote ions (Fe or Mn) with magnetic moments along $+\boldsymbol{\chi}$ and $-\boldsymbol{\chi}$ directions, respectively. In panels (a) and (b), the boxes enclosed by the black dashed lines represent the cells of FeOOH and NaMnP employed in our simulations. The primitive cell of NaMnP is sketched by the box with pink solid lines in panel (b).} 
\end{figure}

Following the analysis in Section V of SM, the FeOOH material has an SPG of $^{\bar{1}}m^{1}m^{1}m^{\infty m}1$. According to Table~\ref{tab:ME}, the $^{\bar{1}}m^{1}m^{1}m^{\infty m}1$ SPG enables linear magnetoelectric NRZSSs via the $E_x\sigma_\chi$ coupling. The SPG of NaMnP is $^{\bar{1}}4/^{\bar{1}}m^{\bar{1}}m^{1}m^{\infty m}1$, and this group hosts the piezomagnetoelectric NRZSSs induced by the combination of $\eta_{xy}$ and $E_z$ (see the $\eta_{xy}E_z\sigma_\chi$ coupling in Table~\ref{tab:ETAME}).
To validate our analysis, we compute the band structures for FeOOH and NaMnP antiferromagnets. First of all, we neglect the SOI and carry out collinear magnetism calculations. This treats the magnetic moments of Fe in FeOOH and Mn in NaMnP as scalar quantities (see Fig.~\ref{fig:stru}). On this condition, the electronic spin degree of freedom appears as unmixed spin-up $+s$ and spin-down $-s$ states, where $\chi$ in $s_\chi$ is omitted at the collinear magnetism level.

\begin{figure}[!ht]
\includegraphics[width=1.\linewidth]{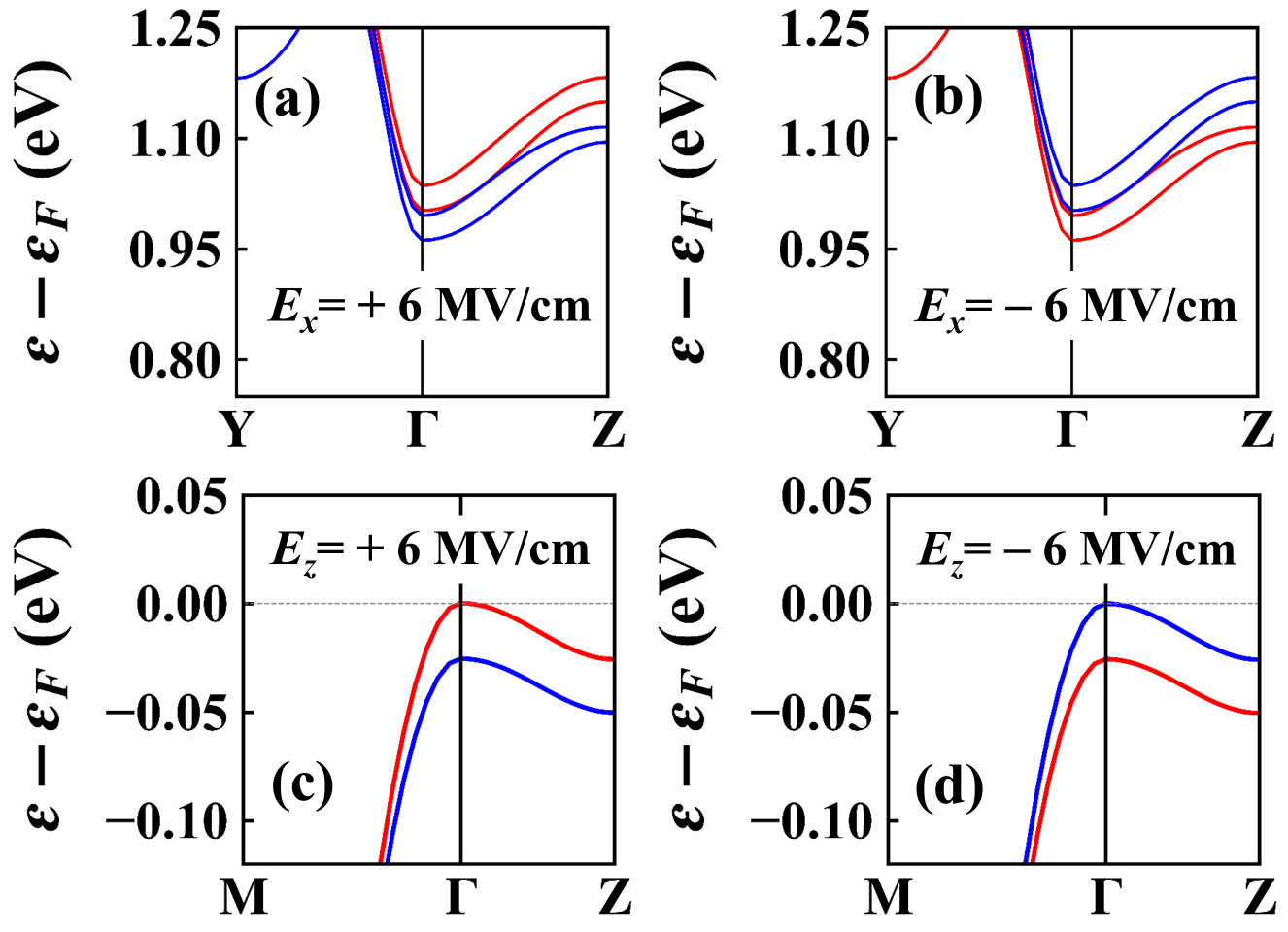}
\caption{\label{fig:band} The electric-field switchable NRZSSs in FeOOH and NaMnP. Panels (a) and (b) are magnetoelectric NRZSSs in FeOOH driven by $E_x$, while panels (c) and (d) are piezomagnetoelectric NRZSSs in NaMnP driven by $E_z$ combined with $\eta_{xy}=4\%$. The red (blue) solid lines correspond to the spin-up (spin-down) states, and $\varepsilon_F$ denotes the Fermi level.} 
\end{figure}

\begin{figure}[!ht]
\includegraphics[width=0.8\linewidth]{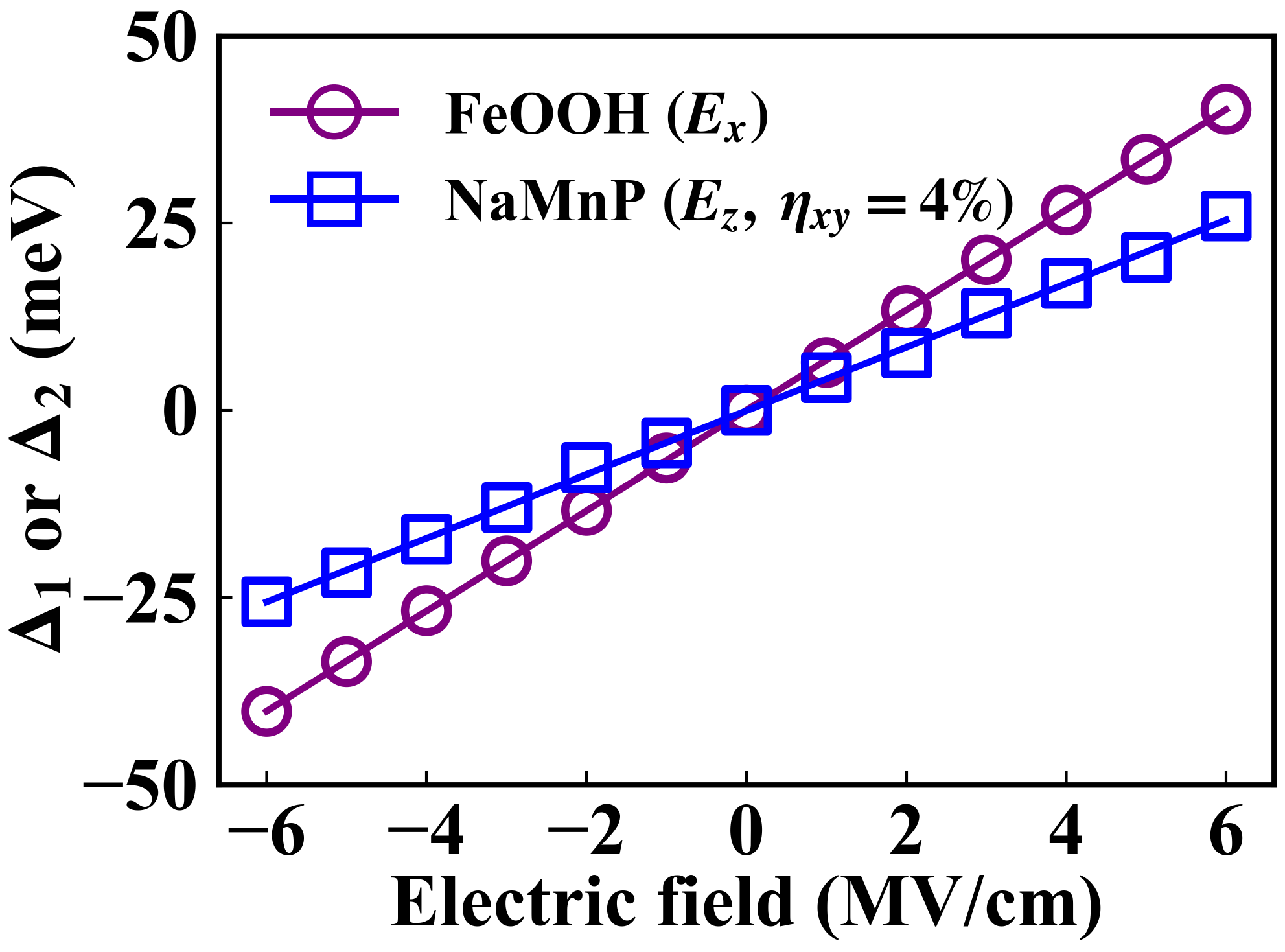}
\caption{\label{fig:NRZSSs} The linear magnetoelectric $\Delta_1$ splittings in FeOOH and bilinear piezomagnetoelectric $\Delta_2$ splittings in NaMnP. Purple circles show $\Delta_1$ in FeOOH as a function of $E_x$; Blue squares show $\Delta_2$ in NaMnP as a function of $E_z$ (under a fixed strain of $\eta_{xy}=4\%$). Purple and blue lines display the corresponding linear regression results.} 
\end{figure}

As shown in Fig.~S2 of the SM, both the conduction band minimum (CBM) of FeOOH and the valence band maximum (VBM) of NaMnP locate at the Brillouin zone center. We shall focus on such two band edges in our following discussion. Under null electric field, the spin-up $\varepsilon_\mathrm{C}(+s)$ and spin-down $\varepsilon_\mathrm{C}(-s)$ energy levels of FeOOH are degenerate at the CBM. Applying an electric field along the $x$ direction splits the $\varepsilon_\mathrm{C}(+s)$ and $\varepsilon_\mathrm{C}(-s)$ energy levels. To be accurate, the $\Delta_1=\varepsilon_\mathrm{C}(+s)-\varepsilon_\mathrm{C}(-s)$ splittings are $\pm  40$ meV for $E_x=\pm 6$ MV/cm [see Figs.~\ref{fig:band}(a) and~\ref{fig:band}(b)]. For unstrained NaMnP under null electric field, the spin-up $\varepsilon_\mathrm{V}(+s)$ and spin-down $\varepsilon_\mathrm{V}(-s)$ energy levels at the VBM are degenerated as well. In the presence of epitaxial strain $\eta_{xy}=4\%$, the $\Delta_2=\varepsilon_\mathrm{V}(+s)-\varepsilon_\mathrm{V}(-s)$ splittings of $\pm 26$ meV can be created by applying $E_z$ of $\pm 6$ MV/cm. Figure~\ref{fig:NRZSSs} shows the $\Delta_1$ splittings in FeOOH as a function of $E_x$ and the $\Delta_2$ splittings in NaMnP as a function of $E_z$ ($\eta_{xy}=4\%$). This confirms that both the magnetoelectric NRZSSs in FeOOH (CBM) and piezomagnetoelectric NRZSSs (VBM) in NaMnP depend linearly on the corresponding electric fields, and are electrically switchable.

To complete this section, we check the relativistic effect on the electric-field induced Zeeman-type spin splittings in FeOOH and NaMnP. We calculate the band structures for FeOOH and NaMnP at noncollinear magnetism level (with SOI). This treats the magnetic moments of Fe and Mn as vectors along $\pm\boldsymbol{\chi}$ directions ($\boldsymbol{\chi}=\mathbf{x},\mathbf{y},\mathbf{z}$). Because of the SOI, the spin-up and spin-down eigen states are not well defined in FeOOH and NaMnP. Instead, the electronic spin degree of freedom is characterized by the spin angular momentum vector, whose predominant component is along $+\boldsymbol{\chi}$ or $-\boldsymbol{\chi}$ direction. In this case, we may use $\Delta^\prime_1=\varepsilon_\mathrm{C}(+s_\chi)-\varepsilon_\mathrm{C}(-s_\chi)$ and $\Delta^\prime_2=\varepsilon_\mathrm{V}(+s_\chi)-\varepsilon_\mathrm{V}(-s_\chi)$ to describe the Zeeman-type spin splittings associated with the CBM of FeOOH and the VBM of NaMnP, respectively (see the previous paragraph for the definitions of $\Delta_1$ and $\Delta_2$). In Fig.~S3 of the SM, we show the band structures with Zeeman-type spin splittings for FeOOH and NaMnP, calculated at the relativistic level. Basically, the inclusion of relativistic corrections does not affect the magnitude of the Zeeman-type spin splittings in FeOOH and NaMnP.\\

\noindent
\textit{Summary and outlook ---} By analyzing the symmetries of spin point groups, we develop a theory that describes the NRZSSs in collinear antiferromagnets driven by electric field. In particular, we identify the linear magnetoelectric and bilinear piezomagnetoelectric mechanisms, both of which are associated with NRZSSs depending linearly on the applied electric field. The reversal of electric field flips the electronic spin states, yielding the electric-field switchable NRZSSs. Our central results are summarized in Tables~\ref{tab:ME} and~\ref{tab:ETAME}, and this guides the first-principles predictions of linear magnetoelectric NRZSSs in FeOOH and bilinear piezomagnetoelectric NRZSSs in NaMnP.

Our theory provides guidelines for the discovery of light-element collinear antiferromagnets with linear magnetoelectric or bilinear piezomagnetoelectric NRZSSs. The NRZSSs therein resemble Zeeman spin splittings in conventional ferromagnets or ferrimagnets, but are more attractive due to the merits of antiferromagnets (e.g., ultralow stray field and fast spin dynamics~\cite{baltz2018antiferromagnetic,rimmler2024non}). Moreover, light-element materials exhibit weak SOIs that result in suppressed spin relaxations and enhanced spin life times, and the electrically switchable NRZSSs enable the manipulation of spin in an energy-efficient electrical manner. These features highlight the potential of such light-element collinear antiferromagnets in designing high-performance spin-based information devices.\\

\noindent
\textit{Acknowledgements ---} The authors acknowledge the support from the National Natural Science Foundation of China (Grants No.~12274174, No.~52288102, No.~52090024, and No.~12034009). L.B. thanks the Vannevar Bush Faculty Fellowship (VBFF) grant No. N00014-20-1-2834 from the Department of Defense and award No. DMR-1906383 from the National Science Foundation AMASE-i Program (MonArk NSF Quantum Foundry). L.Y. thanks the support from high-performance computing center of Jilin University. H.J.Z. acknowledges the support from ``Xiaomi YoungScholar'' Project.\\

\input{main.bbl}
\end{document}

%% file: main.bbl
%